# An Elliptic Curve-based Signcryption Scheme with Forward Secrecy [†]

Mohsen Toorani [‡]         Ali A. Beheshti

## Abstract

*An elliptic curve-based signcryption scheme is introduced in this paper that effectively combines the functionalities of digital signature and encryption, and decreases the computational costs and communication overheads in comparison with the traditional signature-then-encryption schemes. It simultaneously provides the attributes of message confidentiality, authentication, integrity, unforgeability, non-repudiation, public verifiability, and forward secrecy of message confidentiality. Since it is based on elliptic curves and can use any fast and secure symmetric algorithm for encrypting messages, it has great advantages to be used for security establishments in store-and-forward applications and when dealing with resource-constrained devices.*

## 1. Introduction

The encryption and digital signature are two fundamental cryptographic mechanisms that can provide the security of communications. Until the before decade, they have been viewed as important but distinct building blocks of various cryptographic systems. In the public key schemes, a traditional method is to digitally sign a message then followed by an encryption (signature-then-encryption) that can have two problems: Low efficiency and high cost of such summation, and the case that any arbitrary scheme cannot guarantee the security. The signcryption is a relatively new cryptographic technique that is supposed to fulfill the functionalities of digital signature and encryption in a single logical step. It effectively decreases the computational costs and communication overheads in comparison with the traditional signature-then-encryption schemes. The first signcryption scheme was introduced by Zheng (1997) but it fails the forward secrecy of message confidentiality (Jung et al. 2001). Zheng also proposed an elliptic curve-based signcryption scheme that saves 58% of computational and 40% of communication costs when compared with the traditional elliptic curve-based signature-then-encryption schemes (Zheng and Imai, 1998). Several signcryption schemes are also proposed over the years, each of them providing different level of security services and computational costs. The correctness, efficiency, and security are the essential attributes that any signcryption scheme should take them into account. A signcryption scheme should simultaneously fulfill the security attributes of an encryption and those of a digital signature. Such properties mainly include: *Confidentiality, Unforgeability, Integrity*, and *Non-repudiation*. Some signcryption schemes provide further attributes such as *Public verifiability* and *Forward secrecy of message confidentiality* while the others do not provide them. The public verifiability may not be required in some applications while forward secrecy of message confidentiality has an increasingly significance especially when the signcryption is to be done on poorly protected devices such as mobile phones. For the case of brevity, we do not describe the mentioned attributes since they can be easily found in the corresponding literature such as (Tso et al., 2007).

Currently, the elliptic curve cryptography is being used in a wide variety of applications. The elliptic curve-based systems can attain to a desired security level with significantly smaller keys than those of required for their counterparts. This can enhance the speed and leads

---

to an efficient use of power, bandwidth, and storage that are the basic limitations of resource-constrained devices. The *Elliptic Curve Discrete Logarithm Problem* (ECDLP) (Hankerson et al., 2004) that can be a computationally infeasible problem has an essential role in the elliptic curve-based approaches.

In this paper, a new elliptic curve-based signcryption scheme is introduced that simultaneously provides the attributes of message confidentiality, authentication, integrity, unforgeability, non-repudiation, public verifiability, and forward secrecy of message confidentiality. It is an authenticated scheme since it deploys an implicit authenticated key establishment.

## 2. The Proposed Scheme

Throughout this paper, *Alice* is the sender, *Bob* is the recipient, and *Mallory* is the malicious active attacker. Our proposed signcryption scheme is depicted in Figure 1 where some of its deployed notations are described in Figure 2. It consists of four phases: Initialization, Signcryption, Unsigncryption, and Judge Verification. The initialization phase includes selecting the domain parameters, generating the private/public keys, and getting a certificate for the public key of each user. In signcryption phase, *Alice* signcrypts her message and sends it to *Bob*. In unsigncryption phase, *Bob* performs the unsigncryption to recover the signcrypted text and verify the signature. The judge verification phase is used only when any dispute occurs in which the judge decides whether *Alice* has sent the signcrypted message to *Bob* or not.

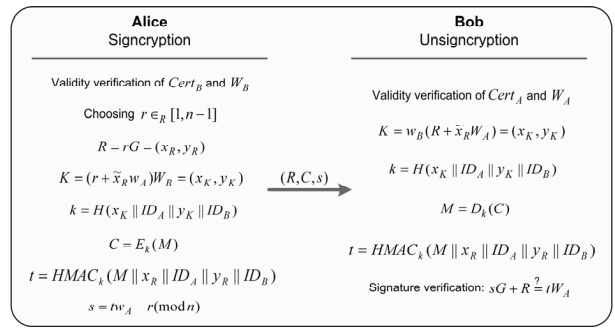

**Fig.1. The proposed signcryption scheme**

| Notations Guide | | | |
|---|---|---|---|
| $\in_R$ | Chosen randomly | $n$ | Order of $G$ ($nG=O$) |
| $M$ | Plaintext | $ID_A$ | Identifier of *Alice* |
| $C$ | Ciphertext | $ID_B$ | Identifier of *Bob* |
| $s$ | Digital signature | $w_A/W_A$ | Private/Public key of *Alice* |
| $H$ | One-way hash function | $w_B/W_B$ | Private/Public key of *Bob* |
| $\|\|$ | Concatenation | $x_R/y_R$ | x/y-coordinate value of point R |
| $G$ | Base point of elliptic curve | $E_k(.)/D_k(.)$ | Symmetric encryption/decryption using the secret key $k$ |
| $O$ | Point of elliptic curve at infinity | | |

**Fig.2. Explanation of deployed notations**

### 2.1. Initialization

Domain parameters of the proposed scheme consist of a suitably selected elliptic curve $E$ defined over a finite field $F_q$ with the Weierstrass equation of the form $y^2 = x^3 + ax + b$, and a *base point* $G \in E(F_q)$ in which $q$ is a large prime number. In order to make the elliptic curve non-singular, $a, b \in F_q$ should satisfy $4a^3 + 27b^2 \neq 0 (\mod q)$ (Hankerson et al., 2004). To guard against *small subgroup attacks*, the point $G$ should be of a prime order $n$, or equivalently $nG = O$, and we should have $n > 4\sqrt{q}$ (Law et al., 2003). To protect against other known attacks on special classes of elliptic curves, $n$ should not divide $q^i - 1$ for all $1 \leq i \leq V$ ($V = 20$ suffices in practice), $n \neq q$ should be satisfied, and the curve should be non-supersingular (Law et al., 2003). In order to keep the intractability of ECDLP to the Pollard-rho and Pohlig-Hellman algorithms (Stinson, 2006), $n$ should at least satisfy $n > 2^{160}$ for the common applications. NIST has also specified the minimum bit lengths of elliptic curves' domain parameters for different required level of security (NIST, 2007).

The private keys of *Alice* and *Bob* are the randomly selected integers $w_A, w_B \in_R [1, n-1]$. The corresponding public keys are calculated as $W_A = w_A G$ and $W_B = w_B G$. *Alice* and *Bob* are uniquely identified by the unique identifiers $ID_A$ and $ID_B$ respectively. They also get the certificates $Cert_A$ and $Cert_B$ from the *Certificate Authority* (*CA*) for their public keys $W_A$ and $W_B$. If *CA* is not involved in the public key generation that is generally the case, it is necessary for *CA* to verify that each entity really possesses the corresponding private key of its claimed public key. This can be accomplished by a zero-knowledge proof. It should also be verified that the public keys belong to the main group. Hereafter, it is also assumed that the participants have access to an authentic copy of the *CA*'s public key, in order to use it for the purpose of certificate validation. The process of certificate validation includes (Stinson, 2006):

(a) Verifying the integrity and authenticity of the certificate by verifying the *CA*'s signature on the certificate.
(b) Verifying that the certificate is not expired.
(c) Verifying that the certificate is not revoked.

### 2.2. Signcryption

*Alice* generates the signcrypted text $(R, C, s)$ by following the below steps:

(1) Checks the validity of $Cert_B$ and uses it for verifying $W_B$.
(2) Randomly selects an integer $r \in_R [1, n-1]$.
(3) Computes $R = rG = (x_R, y_R)$.
(4) Computes $K = (r + \tilde{x}_R w_A)W_B = (x_K, y_K)$ where $\tilde{x}_R = 2^{\lceil f/2 \rceil} + (x_R \bmod 2^{\lceil f/2 \rceil})$ in which $f = \lfloor \log_2 n \rfloor + 1$ is the bit length of *n*, $\lfloor . \rfloor$ denotes the floor, and $\lceil . \rceil$ indicates the ceiling. If $K = O$ she goes back to step 2. Otherwise, she drives the session key of encryption as $k = H(x_K \| ID_A \| y_K \| ID_B)$ in which *H* generates the required number of bits as the secret key of deployed symmetric encryption.
(5) Computes the ciphertext as $C = E_k(M)$.
(6) Computes the digital signature as $s = tw_A - r (\bmod n)$ in which $t = HMAC_k(M \| x_R \| ID_A \| y_R \| ID_B)$ where HMAC is the *Hash Message Authentication Code*.
(7) Sends the signcrypted text $(R, C, s)$ to *Bob*.

### 2.3. Unsigncryption

*Bob* who received the signcrypted text $(R, C, s)$, follows the below steps to extract the plaintext and verify the signature:

(1) Checks the validity of $Cert_A$ and uses it for verifying $W_A$.
(2) Computes $K = w_B(R + \tilde{x}_R W_A) = (x_K, y_K)$ and derives the session key as $k = H(x_K \| ID_A \| y_K \| ID_B)$.
(3) Decrypts the received ciphertext as $M = D_k(C)$.
(4) Computes $t = HMAC_k(M \| x_R \| ID_A \| y_R \| ID_B)$.
(5) Verifies the *Alice*'s signature by verifying $sG + R = tW_A$. If this condition is satisfied, *Bob* accepts *M* as the correct plaintext of *Alice*. Otherwise, he rejects *M*.

In the security policies of certain applications, it may be specified that *Bob* upon verifying the signature of *Alice*, sends her a confirmation message $M''$ in addition to a tag $t'' = MAC_k(M'')$ in which MAC is the Message Authentication Code. In such cases, *Bob* should also verify the validity of the received $R = (x_R, y_R)$, in order to thwart the *invalid-curve attack* (Hankerson et al., 2004). The point *R* acts as an ephemeral public key while *r* is the corresponding ephemeral private key. The validity verification of such ephemeral public key mainly includes checking the following conditions (Antipa et al., 2003):

(a) $R \neq O$,
(b) $x_R, y_R \in F_q$,
(c) *R* should satisfy the defining equation of *E*.

Here, the unsigncryption will be terminated with failure if any of the above-mentioned conditions fails.

### 2.4. Judge Verification

When *Bob* claims that he has received the signcrypted text $(R, C, s)$ from *Alice* and a dispute occurs, the trusted third party (judge) wants *Bob* to provide $(R, C, s, M, k)$. *Bob* is simply capable of extracting *M* and *k* from the previously saved $(R, C, s)$. The judge follows the following steps to adjudicate on what *Bob* claims.

(1) Checks the validity of $Cert_A$ and uses it for verifying $W_A$.
(2) Verifies $M = D_k(C)$. If this is not the case, *Bob* is wrong.
(3) Computes $t = HMAC_k(M \| x_R \| ID_A \| y_R \| ID_B)$.

(4) Verifies the signature of *Alice* by checking the $sG + R = tW_A$ condition. If this condition is not satisfied, *Bob* is wrong. Otherwise, *Alice* has sent $(R, C, s)$ to *Bob*.

Several precautions should be taken into account in implementation of the proposed scheme. It is strongly recommended to use a strong block cipher (such as AES) for encrypting messages. The generated random numbers and also the state of pseudo-random generators should be erased from the memory as soon as the signature is generated. Using the predefined pairs of $(r, R)$ and saving them in an insecure storage media is not recommended. The private keys should be stored in a secure storage media such as a *Hardware Security Module* (HSM).

Although it is not recommended, the first step of the signcryption and unsigncryption phases may be simplified after the first run of the protocol. *Alice* and *Bob* may save the trusted public key of the other party for the future uses. However, it is still necessary for *Bob* to check the revocation status of the *Alice*'s certificate. The revocation status can be obtained using either of *Certificate Revocation List* (CRL), Delta CRL, or *Online Certificate Status Protocol* (OCSP). The OCSP ([Myers et al. 1999](#)) is the most profitable solution for the resource-constrained devices that cannot save a too large CRL, and whenever the timely information is necessary, e.g. in the funds transferring. The OCSP responses are digitally signed with a private key that its corresponding trusted public key is known to the participants. The OCSP server may query its required information from a database server. The certificates are usually stored in an LDAP (Lightweight Directory Access Protocol) directory ([Zeilenga, 2006](#)). Figure 3 depicts a typical configuration for the proposed scheme when an OCSP server is queried for the revocation status, in which $OCSP_A$ and $OCSP_B$ are the corresponding OCSP tokens for the certificates of *Alice* and *Bob*. For the resource-constrained environments, delegating the validity verifications to a trusted server will improve the performance. The verification authority and hash chain can also be used for simplifying the certificate validations ([Satizabál, 2007](#)).

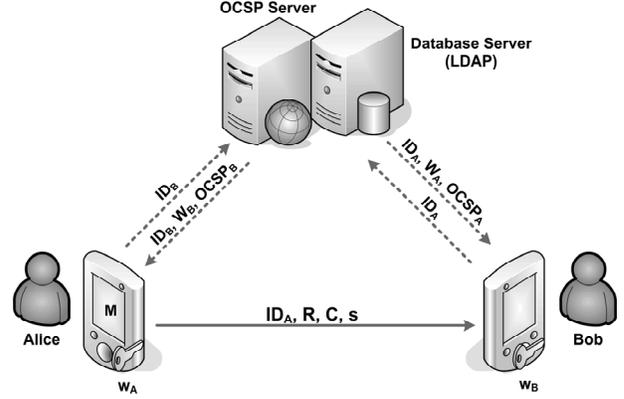

**Fig.3. A typical configuration for the proposed scheme**

The correctness of the proposed scheme can be simply verified. *Alice* and *Bob* will reach to the same point $K$ on the elliptic curve since:

$$K_A = (r + \tilde{x}_R w_A)W_B = (r + \tilde{x}_R w_A)w_B G = \\ = w_B(R + \tilde{x}_R W_A) = K_B = (x_K, y_K) \quad (1)$$

Subsequently, both of participants calculate the same session key as $k = H(x_K \| ID_A \| y_K \| ID_B)$ so *Bob* correctly decrypts the signcrypted text and verifies the signature.

Defining $\tilde{x}_R$ as the least significant half in binary representation of $x_R$ is just a trade-off between security and efficiency. This corresponds with the NIST specifications ([NIST, 2007](#)). Let $c$ denotes the bit length of $\tilde{x}_R$. The smaller the $c$ the less required number of operations will be, and the efficiency will be subsequently improved. However, a too small $c$ may decrease the security. For a similar case, Krawczyk has shown that choosing $c$ so that $2^c > \sqrt{f}$ does not increase the security ([Krawczyk, 2005](#)). If $L$ denotes a point of elliptic curve and $\rho$ is a positive integer, finding $\rho L$ takes $O(\log|\rho|)$ group operations ([Wagstaff, 2003](#)). As an example, the required number of operations for calculating $\tilde{x}_R W_A$ is decreased by a factor of $\frac{\log(2^{2c}-1)}{\log(2^c-1)} \approx \frac{\log(2^{2c})}{\log(2^c)} = 2$ with respect to that is required for calculating $x_R W_A$ when $\tilde{x}_R$ is taken as the half in binary representation of $x_R$. Therefore, setting $c = \lceil f/2 \rceil$ provides the best trade-off between security and efficiency.

## 3. On the Security of the Proposed Scheme

The proposed scheme provides a wide variety of security attributes as it is depicted in Table 1. The long-term private key of *Alice* is involved in the session key generation so the session key has resilience to disclosure of secret value *r*. Validity verification of the static and ephemeral public keys, and the certificates are carefully considered so several kinds of attacks are thwarted. The proposed scheme gets its security from several components:

1) The security attributes of the session key establishment,
2) The security attributes of the certificates,
3) The security attributes of deployed block cipher, one-way hash function, and HMAC,
4) Intractability of ECDLP due to the selected domain parameters.

The proposed scheme deploys a strong key establishment. Until now, many authenticated key exchange protocols are introduced, each of them having their own problems and limitations. The MQV protocols (Law et al., 2003) are possibly the most efficient of all known authenticated Diffie-Hellman protocols that use public-key authentication. The MQV has been widely standardized, and has been selected by the NSA to protect the classified information of USA government. Although HMQV (Krawczyk, 2005) tries to thwart the MQV's vulnerabilities by basically introducing an additional hash function, it also has several vulnerabilities (Menezes, 2005. Menezes and Ustaoglu, 2006). The elliptic curve-based session key establishment process of the proposed scheme does not exactly correspond with that of (Law et al., 2003) and (Krawczyk, 2005), but it tries to improve and match such ideas for its own case. The session key establishment part of the proposed scheme has itself the following security attributes:

1. *Known session key security:* Each execution of the protocol results in a unique session key. The session key will differ for different sessions because the ephemeral random number *r* is introduced in the session key establishment process so the compromise of one session key does not compromise the keys of the other sessions. Since the private keys and identifiers of both participants are involved in the session key derivation function, it will differ even if *Alice* uses the same random number *r* for signcrypting the same message for different recipients.

2. *Resilience to the Unknown-Key Share attack:* In an UKS attack (Kaliski, 2001), two parties compute the same session key but have different views of their peers in the key exchange. This attack is feasible when a key exchange protocol fails to provide an authenticated binding between the session key and identifiers of the honest entities. In the proposed scheme, validity of certificates and also the static and ephemeral public keys are verified. The UKS attack is thwarted because the identifiers of both *Alice* and *Bob* are explicitly involved in the session key derivation function.

**Table 1. The provided attributes of different signcryption schemes**

| Signcryption Schemes | Confidentiality | Integrity | Unforgeability | Non-repudiation | Public Verifiability | Forward Secrecy |
|---|---|---|---|---|---|---|
| (Zheng, 1997) | Yes | Yes | Yes | By Supplementary Protocol | No | No |
| (Jung et al., 2001) | Yes | Yes | Yes | By Supplementary Protocol | No | Yes |
| (Zheng and Imai, 1998) | Yes | Yes | Yes | By Supplementary Protocol | No | No |
| (Bao and Deng, 1998) | Yes | Yes | Yes | Directly | Yes | No |
| (Gamage et al., 1999) | Yes | Yes | Yes | Directly | Yes | No |
| (Han et al., 2004) | No [a] | No [a] | No [a] | Directly | Yes | No |
| (Hwang et al., 2005) | No [b] | No [b] | No [b] | Directly | Yes | No |
| Our Scheme | Yes | Yes | Yes | Directly | Yes | Yes |

[a] See (Toorani and Beheshti Shirazi, 2010)
[b] See (Toorani and Beheshti Shirazi, 2008)

3. *Resilience to the Key Compromise Impersonation (KCI) attack:* In a KCI attack (Strangio, 2006), *Mallory* who could obtain the private key of *Alice* (but does not have the private key of *Bob*) tries to impersonate another honest party *Bob* to *Alice*. Resistance to the KCI attack is an important feature since as long as *Mallory* cannot actively control *Alice*, any session that is established by *Alice* remains secure. Under intractability of the ECDLP, the KCI attack is thwarted in the proposed scheme. An adversary that could obtain $w_A$, should find the corresponding $r$ of $R$ in order to deduce the corresponding session key that is generally in deposit of solving the ECDLP.

4. *Partial Forward secrecy:* A system has the attribute of *partial forward secrecy* if compromising the long-term private key of one entity (or more but not all the entities) does not compromise the previously established session keys. In the proposed scheme, even if $w_A$ is revealed, the attacker should have the value of the corresponding $r$ that is generally in deposit of solving the ECDLP.

As it is indicated in Table 1, the proposed scheme provides many attributes. Hereunder, a brief proof is given for the claimed attributes.

1) *Confidentiality*: The proposed scheme deploys a strong block cipher so according to the *Kerckhoffs'* principle, the secrecy will be entirely resided in the established session keys. The session key establishment process has itself several security attributes. Ultimately, an adversary has only two ways to defeat the confidentiality: having $w_B$, or deriving both $w_A$ and $r$. Deriving the private keys and finding the corresponding $r$ of a specific $R$ is in deposit of solving the ECDLP that is computationally infeasible with the selected domain parameters. If $w_B$ is revealed, the attacker can only decrypt those ciphertexts that are signcrypted by $W_B$. This is an inevitable feature of all the one-pass protocols with implicit authenticated key exchange. However, for a definite sender, the attacker should have both $w_A$ and the corresponding $r$ of the clearly sent $R$. Otherwise, he cannot derive the true session key and the decryption will be computationally infeasible. It is noteworthy that the proposed scheme does not have the feature of *past recovery* so *Alice* cannot retrieve her previously signcrypted texts without having the corresponding $r$ of $R$. Compromising the *past recovery* is a common feature of all the forward secure systems.

2) *Authentication*: The implicit authentication is provided in three ways: the proposed scheme is certificate-based and the certificates are verified by both sender and recipient. An implicit authentication is also involved in the session key establishment so only the correct party who has the true private key can reach to the correct key agreement and perform the unsigncryption. An authentication is also accomplished when *Bob* verifies the signature by checking the $sG + R = tW_A$ condition.

3) *Unforgeability*: *Mallory* cannot forge the valid $(M, R, s)$ with his malicious $(M', R, s')$. A valid forged signature $s'$ should satisfy $t'W_A - s'G = tW_A - sG$ or equivalently $s' = s + (t' - t)w_A$ so the knowledge of $t$, $t'$ and $w_A$ is necessary. To find the true value of $t = HMAC_k(M \| x_R \| ID_A \| y_R \| ID_B)$ and $t' = HMAC_k(M' \| x_R \| ID_A \| y_R \| ID_B)$, *Mallory* should know the session key $k$ that is in deposit of his knowledge of $w_B$, or both $w_A$ and $r$. Even if he knows $w_B$, he can only find the values of $t$ and $t'$ but his knowledge of $w_A$ is still necessary to forge the signature as $s' = s + (t' - t)w_A$. Otherwise, he cannot truly forge the signature and his forged signature will be recognized in unsigncryption and judge verification phases when checking the $sG + R = tW_A$ condition.

4) *Non-repudiation*: This attribute can be deduced from the unforgeability. It is computationally infeasible to forge the *Alice*'s signature without having $w_A$. The signature is verified through the certificate and public keys validation and when checking the $sG + R = tW_A$ condition.

5) *Integrity*: The hash value of message concatenated with some variable parameters is involved in the signature generation so any message alteration will change the signature. The integrity is guaranteed by the security attributes of HMAC and unforgeability of the signature. *Mallory* should also have the valid session key to encrypt his modified message. Otherwise, his modified message will not be correctly decrypted by *Bob*. The session key is also involved in signature generation. He should have $w_B$, or both $w_A$ and $r$ to find the valid session key. He should also find a collision for the HMAC that is computationally infeasible. Otherwise, he cannot forge a valid signature. The integrity is implicitly verified when *Bob* checks the $sG + R = tW_A$ condition.

6) *Public verifiability*: Given $(R, C, s, M, k)$ anybody can verify the signature by checking the $sG + R = tW_A$ condition, without any need for the private key of *Alice* or *Bob*. However, in some cases, it may be desirable to keep the message confidentiality so that any third party can verify the sender's signature without any need for knowledge of plaintext and the corresponding session key. This can be referred to as the directly public verifiability. If necessary, the proposed scheme can be modified so that the signature can be directly verified by anyone who observes the transmitted pairs of $(R, C, s)$. Since directly public verification is not required in many applications, it will be introduced as an optional variant in Section 5.

7) *Forward secrecy of message confidentiality*: It means that even if the long-term private key of the sender is revealed, the adversary is not capable of decrypting the previously signcrypted texts. The only way to defeat the message confidentiality is to have $w_B$, or both $w_A$ and $r$. When $w_A$ is revealed, for decryption to be possible, it is necessary to have the value of random number $r$ for the corresponding session. Otherwise, the ephemeral session key cannot be recovered. Generally, the forward secrecy will be compromised only if the attacker can solve the ECDLP that is computationally infeasible with the selected domain parameters. As a one-pass protocol, since there is not any session-specific input from *Bob*, we cannot prospect the proposed scheme for the *Perfect Forward Secrecy* (PFS). However, the proposed scheme provides the partial forward secrecy under intractability of the ECDLP.

## 4. Computational Costs

In this section, the time complexity of the proposed scheme is evaluated. Table 2 gives a comparison between the computational costs of the proposed scheme and those of the other schemes, in which the computational costs of verifications and symmetric encryption are neglected. Since different schemes provide different number of security attributes, it is necessary to take a glimpse at the provided security attributes in Table 1 before comparing the computational costs of different schemes in Table 2. In Table 2, the computational costs are categorized with respect to different kinds of required operations, each of them requiring a different number of bit operations. The total running time of a signcryption scheme depends on the efficiency of deployed algorithm for each of required operations (hardware-independent), and the speed of deployed hardware (hardware-dependent).

Let $\zeta = \lfloor \log_2 n \rfloor + 1$ denotes the bit-length of the modulus $n$. Since $n$ is an odd prime number, the resulted calculations in the modular arithmetic have at most a bit-length of $\zeta$. Adding and subtracting two $\zeta$-bit integers each takes $O(\zeta)$ bit operations (Rosen, 1988) so we have:

$$T_{Add} = O(\zeta) \qquad (2)$$

Multiplying two $\zeta$-bit integers, and dividing a $2\zeta$-bit dividend by a $\zeta$-bit divisor takes $O(\zeta^2)$ bit operations using the conventional method (Rosen, 1988). Thus,

$$T_{Mul} = O(\zeta^2) \qquad (3)$$

$$T_{Div} = O(\zeta^2) \qquad (4)$$

However, one can multiply and divide faster than what is specified by the conventional methods, e.g. a divide-and-conquer algorithm due to Karatsuba and Ofman reduces the complexity of multiplication to $O(\zeta^{\log_2 3})$ (Hankerson et al., 2004). There is also a fast algorithm for multiplying two $\zeta$-bit integers in $O(\zeta \log \zeta \log \log \zeta)$ bit operations but it is not so useful unless $\zeta$ is very large (Wagstaff, 2003). However, for having a fair comparison between the computational costs of different schemes, it is reasonable to just consider what is required by the conventional methods since such results are the results of the worst situation and anyone can decrease the required number of bit operations by deploying faster algorithms. For the modular inverse calculation using the conventional method, we have (Rosen, 1988):

$$T_{Inv} = O(\zeta^3) \quad (5)$$

**Table 2. Computational costs of different signcryption schemes**

| Signcryption schemes | Participant | Modular Exponentiation (Exp) | Modular Division/inverse (Div) | Elliptic Curve Point Multiplication (ECPM) | Elliptic Curve Point Addition (ECPA) | Modular Multiplication (Mul) | Modular Addition (Add) | One-way or keyed one-way hash function (Hash) |
|---|---|---|---|---|---|---|---|---|
| (Zheng, 1997) | Alice | 1 | 1 | - | - | - | 1 | 2 |
| | Bob | 2 | - | - | - | 2 | - | 2 |
| (Jung et al., 2001) | Alice | 2 | 1 | - | - | - | 1 | 2 |
| | Bob | 3 | - | - | - | 1 | - | 2 |
| (Bao and Deng, 1998) | Alice | 2 | 1 | - | - | - | 1 | 3 |
| | Bob | 3 | - | - | - | 1 | - | 3 |
| (Gamage et al., 1999) | Alice | 2 | 1 | - | - | - | 1 | 2 |
| | Bob | 3 | - | - | - | 1 | - | 2 |
| (Zheng and Imai, 1998) | Alice | - | 1 | 1 | - | 1 | 1 | 2 |
| | Bob | - | - | 2 | 1 | 2 | - | 2 |
| (Han et al., 2004) | Alice | - | 1 | 2 | - | 2 | 1 | 2 |
| | Bob | - | 1 | 3 | 1 | 2 | - | 2 |
| (Hwang et al., 2005) | Alice | - | - | 2 | - | 1 | 1 | 1 |
| | Bob | - | - | 3 | 1 | - | - | 1 |
| Our Scheme | Alice | - | - | 2 | - | 2 | 2 | 2 |
| | Bob | - | - | 4 | 2 | - | - | 2 |

The running time of the modular exponentiation using the conventional methods will be of (Rosen, 1988):

$$T_{Exp} = O(\zeta^3) \quad (6)$$

The HMAC executes in approximately the same time as its embedded hash function for long messages but its running time depends on the kind of its embedded hash function. Total number of operations for the HMAC-SHA1 calculations (Elkeelany et al., 2002) is as:

$$T_{HMAC-SHA1} = 32 + 1110 \times (2 + n_k) \quad (7)$$

While for the case of HMAC-MD5, it is:

$$T_{HMAC-MD5} = 32 + 744 \times (2 + n_k) \quad (8)$$

In which $n_k = \dfrac{N+k}{512}$ is the number of input blocks to the embedded hash function where $N$ is the bit-length of the total message, and $k$ is the bit-length of extra-appended inner form of the key (Elkeelany et al., 2002).

The computations in elliptic curves depend on many factors including the deployed coordinates. The computational cost of an *Elliptic Curve Point Addition* (ECPA) in an Affine coordinate is as:

$$T_{ECPA|_{Affine}} = 3T_{Mul} + 6T_{Add} + T_{Inv} \quad (9)$$

While for the Jacobian projective coordinate, we have:

$$T_{ECPA|_{Jacobian}} = 16T_{Mul} + 7T_{Add} \quad (10)$$

The *Elliptic Curve Point Multiplication* (ECPM) is a basic and time-consuming operation in ECC. The execution times of the EC-based schemes are typically dominated by the point multiplication. It typically takes $O(\log|k|)$ group operations to compute $kP$ when $k \neq 0$ (Wagstaff, 2003). Several methods for the scalar multiplication are proposed in the literature. However, selecting the suitable algorithm is complicated by platform characteristics, coordinate selection, memory and other constraints, security considerations, and interoperability requirements (Hankerson et al., 2004).

For elliptic curve P-192 defined over $GF(p = 2^{192} - 2^{64} - 1)$ that provides the same level of security as that of RSA with a 1024-bits modulus (NIST, 2000), the total required number of field operations for a point multiplication for an unknown point using the *Window NAF* method on Jacobian-Chudnovsky coordinate is calculated in (Hankerson et al., 2004) as:

$$T_{ECPM}|_{\text{Jacobian-Chudnovsky}} = 1936 T_{Mul} + T_{Inv} \qquad (11)$$

However, for a fixed point *P*, using offline pre-computations and Comb 2-table method in Jacobian-Affine coordinate, it can be calculated as (Hankerson et al., 2004):

$$T_{ECPM}|_{\text{Jacobian-Affine}} = 638 T_{Mul} + T_{Inv} \qquad (12)$$

that is explicitly less than what is specified in (11). For other curves, coordinates and algorithms, the required number of operations will differ from what is specified in (11) and (12). The computational costs of the proposed scheme and those of the other schemes can be fairly compared using Table 2 and expressions (2-11). Figure 4 depicts the total required number of bit operations for executing signcryption and unsigncryption in the signcryption schemes that are described in Table 2, in which the computational costs of verifications and symmetric encryption are neglected. As it is apparent from Figure 4, the computational costs of our proposed scheme in unsigncryption phase is slightly more than the other depicted EC-based schemes, as it provides the most feasible security attributes, but it has a great computational advantage over its exponentiation-based counterparts.

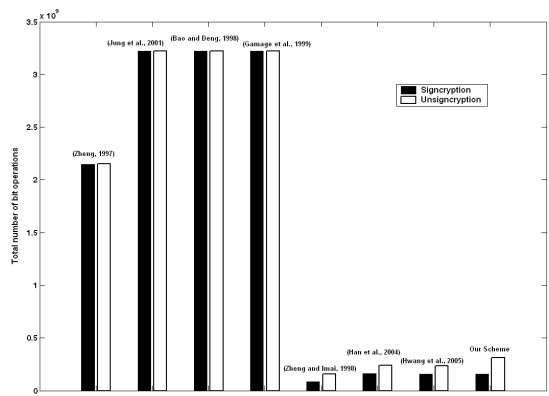

**Fig. 4. Total number of bit operations for different signcryption schemes**

## 5. Optional Improvements

In this section, some improvements are introduced to the proposed scheme. They are considered as optional improvements since they may not be required in some applications and systems.

### 5.1. Delegating verifications to a trusted server

When the participants are dealing with processing and memory constraints, delegating the validity verifications to a trusted server will offer significant advantages. The required time of sending a certificate to a validation server and subsequently, receiving and authenticating the response can be considerably less than the required time of performing the certificate path discovery and validation by a resource-constrained device. The performance of the proposed scheme can be greatly enhanced by delegating the validation processes to a *Delegated Validation* (DV) server, as it is depicted in Figure 5. The DV server can be independent of the communication network. It accomplishes the certificate validation via the *Delegated Path Validation* (DPV) protocol but its duties defers from what is specified in (Pinkas and Housley, 2002). All the signcrypted texts will be directed to the DV server through the communication network. The DV server accomplishes the certificate and public key validations for both sender and recipient. It queries the database server for the certificates of both sender and recipient through their identifiers. It obtains the revocation statuses by getting OCSP responses from the OCSP server. It also checks the validity of point *R*. The DV server will contact the designated recipient after a successful validation. If any error occurs, the DV server will send an error message to the initiator and saves a copy in its log file. According to the security policies, all the transmitted messages may be separately saved by the DV server for the possible disputes. The DV server digitally signs its responses, unless an error is occurred. The signed responses should include a hash value of all the transmitted parameters in addition to the identifiers of both sender and recipient.

In the optimized configuration, *Bob* will not check the validity of point *R* because it is validated by the DV server. He just checks the DV server's signature. *Alice* does not need to check the revocation status of the *Bob*'s certificate. She just needs to have the public key of *Bob* that can be simply queried from a database server through the *Bob*'s identifier when *Alice* does not have it. She can then save the public key of *Bob* for the future uses.

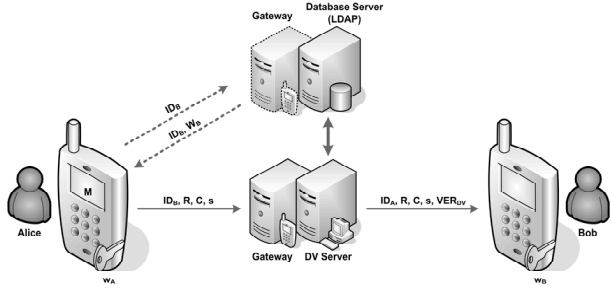

**Fig.5. An optimized configuration for the proposed scheme when it is used for the resource-constrained devices**

## 5.2. Using Timestamps

Timestamps have an unavoidable role in many security applications when the true order of occurrence and freshness of messages are important. *Alice* can produce her timestamp $T_A$ according to her local time or a predefined international time zone. It can be obtained from a trusted time server. The only required modifications to the proposed scheme are to modify the session key derivation function as $k = H(x_K \| ID_A \| y_K \| ID_B \| T_A)$, and parameter $t$ of signature generation function as $t = HMAC_k(M \| x_R \| ID_A \| y_R \| ID_B \| T_A)$. *Alice* clearly includes her timestamp in her signcrypted text as $(T_A, R, C, s)$ and sends it to *Bob*. *Bob* upon receiving a new message, produces his timestamp $T_B$ and checks whether $0 < T_B - T_A < \omega$ in which $\omega$ is the acceptable time window that is determined by estimating the usual delay time between sending and receiving of a message in the deployed communication link. Hereby, *Bob* can easily neglect the old messages. This can simply thwart the potential *replay attacks*. Using timestamps, the deployed encryption and signature generation functions will be time-dependent and the freshness of messages will be enhanced. The timestamp can also be used as a parameter in the seed of random number generators. However, exploiting timestamps has its own problems (e.g. the synchronization) and cannot be deployed in some systems.

## 5.3. Towards directly public verification

In the proposed scheme, the signature can be verified by any third party provided that $(R, C, s, M, k)$ is provided by either of participants (*Alice* or *Bob*). However, in some cases, it may be desirable to keep the message confidentiality so that any third party can verify the sender's signature without any need to the knowledge of plaintext and the corresponding session key. If necessary, the proposed scheme can be modified so that the signature can be directly verified by anyone who observes the transmitted pairs of $(R, C, s)$ and without any for the private keys of the participants. This can be simply accomplished by modifying the parameter $t$ in all phases of the proposed scheme as $t = H(C \| x_R \| ID_A \| y_R \| ID_B)$. Consequently, anyone who observes the transmitted $(R, C, s)$ can compute $t$ and directly verify the *Alice*'s signature by following the below steps:

(1) Check the validity of $Cert_A$ and use it for verifying $W_A$.

(2) Verify the *Alice*'s signature by checking the $sG + R = tW_A$ condition.

## 6. Conclusions

In this paper, a new elliptic curve-based signcryption scheme is introduced that simultaneously provides the security attributes of message confidentiality, authentication, integrity, unforgeability, and non-repudiation. It also has the attribute of public verifiability so any third party can verify the signature without any need for the private keys of the participants. It also has the attribute of forward secrecy of message confidentiality so even if the sender's private key is revealed, no one else can extract the plaintext of the previously signcrypted texts. Since it is based on elliptic curve cryptography and uses symmetric ciphering for encrypting messages, it has great advantages to be deployed in resource-constrained devices such as mobile phones. As a one-pass scheme, it is also so attractive for security establishment in store-and-forward applications such as E-mail and Short Message Service.

# References

Antipa, A., D. Brown, A. Menezes, R. Struik and S. Vanstone, 2003. Validation of elliptic curve public keys. In Proceedings of the 6$^{th}$ International Workshop on theory and Practice in Public Key Cryptography: Public Key Cryptography (PKC'03), London, UK, 6-8 January 2003. LNCS 2567, Springer-Verlag, Berlin/Heidelberg, pp.211-223. DOI: 10.1007/3-540-36288-6_16. Available at: http://www.springerlink.com/content/g9l2fjkdhwbp37b3

Bao, F. and R.H. Deng, 1998. A signcryption scheme with signature directly verifiable by public key. In Proceedings of Advances in Cryptology - PKC'98, LNCS 1431, Springer-Verlag, Berlin, 1998, pp.55-59. DOI: 10.1007/BFb0054009. Available at: http://www.springerlink.com/content/j6fbnefp3eqvp78q

Elkeelany, O., M.M. Matalgah, K.P. Sheikh, M. Thaker, G. Chaudhry, D. Medhi, and J. Qaddour, 2002. Performance Analysis of IPSec Protocol: Encryption and Authentication. In Proceedings of 2002 IEEE International Conference on Communications (ICC'02), Vol.2, pp.1164-1168, 28 April - 2 May 2002, New York City, USA. DOI: 10.1109/ICC.2002.997033. Available at: http://ieeexplore.ieee.org/xpl/freeabs_all.jsp?arnumber=997033

Gamage, C., J. Leiwo and Y. Zheng, 1999. Encrypted message authentication by firewalls. In Proceedings of International Workshop on Practice and Theory in Public Key Cryptography (PKC-99), March 1999. Springer-Verlag, Berlin, LNCS 1560, pp.69-81. DOI: 10.1007/3-540-49162-7_6. Available at: http://www.springerlink.com/content/updpmmnukbytvhvh

Han, Y., X. Yang and Y. Hu, 2004. Signcryption Based on Elliptic Curve and Its Multi-Party Schemes. 3$^{rd}$ ACM International Conference on Information Security (InfoSecu'04), 14-16 November 2004, Shanghai, China. ACM International Conference Proceeding Series, Vol.85, pp.216-217, NY, USA, 2004. DOI: 10.1145/1046290.1046336. Available at: http://portal.acm.org/citation.cfm?doid=1046290.1046336

Hankerson, D., A. Menezes and S. Vanstone (2004). Guide to Elliptic Curve Cryptography, 1$^{st}$ edition. Springer-Verlag, New York. ISBN: 0-387-95273-X.

Hwang, R.-J., C.-H. Lai and F.-F. Su, 2005. An efficient signcryption scheme with forward secrecy based on elliptic curve. Journal of Applied Mathematics and Computation (Elsevier Inc.), 167 (2): 870-881, 2005. DOI: 10.1016/j.amc.2004.06.124. Available at: http://dx.doi.org/10.1016/j.amc.2004.06.124

Jung, H.Y., K.S. Chang, D.H. Lee and J.I. Lim, 2001. Signcryption schemes with forward secrecy. In Proceeding of Information Security Application-WISA 2001, Seoul, Korea, 13-14 September 2001, pp.403-475.

Kaliski, B., 2001. An unknown key-share attack on the MQV key agreement protocol. ACM Transactions on Information and System Security (TISSEC), 4 (3): 275-288, August 2001, NY, USA. DOI: 10.1145/501978.501981. Available at: http://portal.acm.org/citation.cfm?id=501978.501981

Krawczyk, H., 2005. HMQV: A high-performance secure Diffie-Hellman protocol. In Proceedings of Advances in Cryptology – CRYPTO'05. Springer-Verlag, Berlin, LNCS 3621, Nov. 2005, pp.546-566. DOI: 10.1007/11535218_33. Available at: http://www.springerlink.com/content/12qjyqphqb3mge2n

Law, L., A. Menezes, M. Qu, J. Solinas and S. Vanstone, 2003. An efficient Protocol for Authenticated Key Agreement. Journal of Designs, Codes and Cryptography, 28:119-134, March 2003. DOI: 10.1023/A:1022595222606. Available at: http://www.springerlink.com/content/n84gx0461447n518

Menezes, A., 2005. Another Look at HMQV. Nov. 2005. Available at: http://eprint.iacr.org/2005/205.pdf

Menezes, A. and B. Ustaoglu, 2006. On the Importance of Public-Key Validation in the MQV and HMQV Key Agreement Protocols. 7$^{th}$ International Conference on Cryptology in India, Kolkata, India, 11-13 December 2006. In Proceedings of Advances in Cryptology–INDOCRYPT'06, Springer-Verlag, Berlin, LNCS 4329, 2006, pp.133-147. DOI: 10.1007/11941378_11. Available at: http://www.springerlink.com/content/r0kw6t87h7301h53

Myers, M., R. Ankney, A. Malpani, S. Galperin and C. Adams, 1999. X.509 Internet Public Key Infrastructure Online Certificate Status Protocol – OCSP. RFC 2560, June 1999. Available at: http://www.ietf.org/rfc/rfc2560.txt

NIST (National Institute of Standards and Technology), 2007. Special Publication 800-56A, Recommendation for Pair-Wise Key Establishment Schemes Using Discrete Logarithm Cryptography. March 2007. Available at: http://csrc.nist.gov/publications/nistpubs/800-56A/SP800-56A_Revision1_Mar08-2007.pdf

NIST (National Institute of Standards and Technology), 2000. Digital Signature Standard. Federal Information Processing Standards Publication (FIPS) 186-2. January 2000. Available at: http://csrc.nist.gov/publications/fips/fips186-2/fips186-2-change1.pdf

Pinkas, D. and R. Housley, 2002. Delegated Path Validation and Delegated Path Discovery Protocol Requirements. RFC 3379, Sep. 2002. Available at: http://www.ietf.org/rfc/rfc3379.txt

Rosen, K.H. (1988). Elementary Number Theory and Its Applications. 2nd edition, Addison-Wesley, Massachusetts. ISBN: 0201119587.

Satizabál, C., R. Martínez-Peláez, J. Forné and F. Rico-Novella, 2007. Reducing the Computational Cost of Certification Path Validation in Mobile Payment. In Proceedings of Advances in Cryptology–EUROPKI'07, Springer-Verlag, Berlin, LNCS 4582, June 2007, pp.280-296. DOI: 10.1007/978-3-540-73408-6_20. Available at: http://www.springerlink.com/content/g48j504452wv8852

Strangio, M.A., 2006. On the Resilience of Key Agreement Protocols to Key Compromise Impersonation. Advances in Cryptology–EUROPKI'06, Springer-Verlag, Berlin, LNCS 4043, 2006, pp.233-247. DOI: 10.1007/11774716_19. Available at: http://www.springerlink.com/content/t7871tp63r9q4089

Stinson, D.R. (2006). Cryptography-Theory and Practice. 3rd edition. Chapman & Hall/CRC. ISBN: 1-58488-508-4.

Toorani, M. and A.A. Beheshti Shirazi, 2008. Cryptanalysis of an efficient signcryption scheme with forward secrecy based on elliptic curve. Proceedings of 2008 International Conference on Computer and Electrical Engineering (ICCEE'08), 20-22 December 2008, Phuket, Thailand, pp.428-432.

Toorani, M. and A.A. Beheshti Shirazi, 2010. Cryptanalysis of an elliptic curve-based signcryption scheme. International Journal of Network Security, Vol.10, No.1, Jan 2010, pp.51-56.

Tso, R., T. Okamoto and E. Okamoto, 2007. An Improved Signcryption Scheme and Its Variation. In Proceedings of 4th IEEE International Conference on Information Technology (ITNG'07), April 2007, pp.772-778. DOI: 10.1109/ITNG.2007.34. Available at: http://ieeexplore.ieee.org/xpl/freeabs_all.jsp?arnumber=4151775

Wagstaff, S.S. (2003). Cryptanalysis of Number Theoretic Ciphers, 1st edition. Chapman & Hall/CRC. ISBN: 1-58488-153-4.

Zeilenga, K., Lightweight Directory Access Protocol (LDAP): Schema Definitions for X.509 Certificates. RFC 4523, June 2006. Available at: http://www.ietf.org/rfc/rfc4523.txt

Zheng, Y., 1997. Digital signcryption or how to achieve Cost (Signature & Encryption) << Cost (Signature) + Cost (Encryption). In Proceedings of the 17th International Conference on Advances in Cryptology–CRYPTO'97, 17-21 August 1997. Springer-Verlag, Berlin, LNCS 1294, 1997, pp.165-179. DOI: 10.1007/BFb0052234. Available at: http://www.springerlink.com/content/y32t695t4528680k

Zheng, Y. and H. Imai, 1998. How to construct efficient signcryption schemes on elliptic curves. Information Processing Letters, 68: 227-233, Dec. 1998. Elsevier Inc., Amsterdam, Netherlands. DOI: 10.1016/S0020-0190(98)00167-7. Available at: http://portal.acm.org/citation.cfm?id=306049.306053